\documentstyle[psfig]{article}

\def\linebreak{\unskip\break} 

\def\lcdm{$\Lambda$CDM}
\def\gsim{\mathrel{\raise.3ex\hbox{$>$\kern-.75em\lower1ex\hbox{$\sim$}}}}
\def\lsim{\mathrel{\raise.3ex\hbox{$<$\kern-.75em\lower1ex\hbox{$\sim$}}}}

\def\M10{{\times 10^{10} M_{\odot}\ }}

\def\hMpc{\ h^{-1}\ {\rm Mpc}}

\def\kms{\ {\rm km}\ {\rm s}^{-1}}
\def\kmsMpc{\ {\rm km}\ {\rm s}^{-1}\ {\rm Mpc}^{-1}}

\begin{document}

\title{The Best Theory of Cosmic Structure Formation \\
 is Cold + Hot Dark Matter (CHDM)
\footnote{To appear in {\it Critical Dialogues in Cosmology}, 
ed. N. Turok (World Scientific, 1996).}
}

\author{Joel R. Primack\\
University of California, Santa Cruz, CA 95064\\
{\it joel@physics.ucsc.edu}}

\maketitle

\section{Summary}

The fact that the simplest modern cosmological theory, standard Cold
Dark Matter (sCDM), almost fits all available data has encouraged the
search for variants of CDM that can do better.  I have been asked to
make the case here that CHDM is the best theory of cosmic structure
formation, and indeed I believe that it {\it is} the best of all those
I have considered {\it if} the cosmological matter density is near
critical (i.e., $\Omega_0 \approx 1$) and if the expansion rate is not
too large (i.e. $h \equiv H_0/(100 \kmsMpc) \lsim 0.6$).  But I think
it will be helpful to discuss CHDM together with its chief competitor
among CDM variants, low-$\Omega_0$ CDM with a cosmological constant
(\lcdm).  While the predictions of COBE-normalized CHDM and \lcdm\
both agree reasonably well with the available data on scales of $\sim
10$ to $100 \hMpc$, each has potential virtues and defects. \lcdm\
with $\Omega_0 \sim 0.3$ has the possible virtue of allowing a higher
expansion rate $H_0$ for a given cosmic age $t_0$, but the defect of
 predicting too much fluctuation power on small scales.  CHDM has less
power on small scales, so its predictions appear to be in good
agreement with data on the galaxy distribution, although it remains to
be seen whether it predicts early enough galaxy formation to be
compatible with the latest high-redshift data. Also, several sorts of
data suggest that neutrinos have nonzero mass, and the variant of CHDM
favored by this data --- in which the neutrino mass is shared between
two species of neutrinos --- also seems more compatible with the
large-scale structure data.  Except for the $H_0-t_0$ problem, there
is not a shred of evidence in favor of a nonzero cosmological
constant, only increasingly stringent upper bounds on it from several
sorts of measurements. Two recent observational results particularly
favor high cosmic density, and thus favor $\Omega=1$ models such as
CHDM over \lcdm\ --- (1) the positive deceleration parameter $q_0>0$
measured using high-redshift Type Ia supernovae, and (2) the low
primordial deuterium/hydrogen ratio measured in two different quasar
absorption spectra. If confirmed, (1) means that the cosmological
constant probably cannot be large enough to help significantly with
the $H_0-t_0$ problem; while (2) suggests that the baryonic
cosmological density is at the upper end of the range allowed by Big
Bang Nucleosynthesis, perhaps high enough to convert the ``cluster
baryon crisis'' for $\Omega=1$ models into a crisis for low-$\Omega_0$
models. I also briefly compare CHDM to other CDM variants such as Warm
Dark Matter (WDM) and tilted CDM. CHDM has the advantage among
$\Omega=1$ CDM-type models of requiring little or no tilt, which
appears to be an advantage in fitting recent small-angle cosmic
microwave background anisotropy data.  The presence of a hot component
that clusters less than cold dark matter lowers the effective
$\Omega_0$ that would be measured on small scales, which appears to be
in accord with observations, and it may also avoid the discrepancy
between the high central density of dark matter halos from CDM
simulations compared to evidence from rotation curves of dwarf spiral
galaxies.

\section{Introduction}

``Standard'' $\Omega=1$ Cold Dark Matter (sCDM) with $h \approx 0.5$
and a near-Zel'dovich spectrum of primordial fluctuations~\cite{BFPR}
until a few years ago seemed to many theorists to be the most
attractive of all modern cosmological models. But although sCDM
normalized to COBE nicely fits the amplitude of the large-scale flows
of galaxies measured with galaxy peculiar velocity data
\cite{Dekel94}, it does not fit the data on smaller scales: it
predicts far too many clusters \cite{WEF93} and does not account for
their large-scale correlations \cite{Olivier93}, and the shape of the
power spectrum $P(k)$ is wrong \cite{BaughEf94,Zaroubi96}. Here I
discuss what are perhaps the two most popular variants of sCDM that
might agree with all the data: CHDM and \lcdm.  The linear {\it matter}
power spectra for these two models are compared in Figure~1 
(from Ref. \cite{UCLA96}) with the
real-space {\it galaxy} power spectrum obtained from the
two-dimensional APM galaxy power spectrum~\cite{BaughEf94}. The \lcdm\
and CHDM models essentially bracket the range of power spectra in
currently popular cosmological models which are variants of CDM.

\begin{figure}[htb]   
\vskip-1pc
\centering
\centerline{\psfig{file=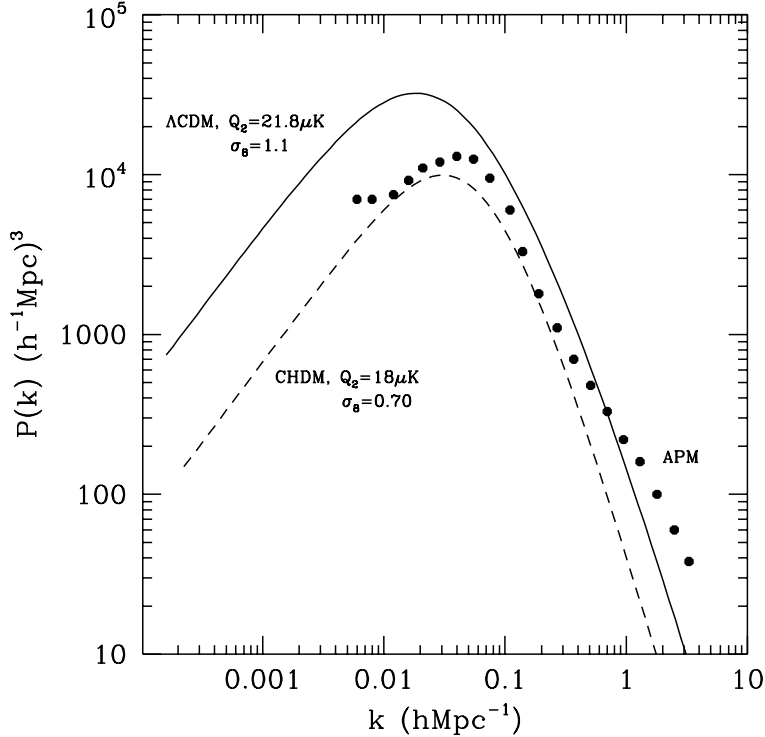,width=10cm}}
\vskip-1pc
\caption{%
Power spectrum of dark matter for \lcdm\ and CHDM models considered in
this paper, both normalized to COBE, compared to the APM galaxy
real-space power spectrum. (\lcdm: $\Omega_0=0.3$,
$\Omega_\Lambda=0.7$, $h=0.7$, thus $t_0=13.4$ Gy; CHDM: $\Omega=1$,
$\Omega_\nu=0.2$ in $N_\nu=2$ $\nu$ species, $h=0.5$, thus $t_0=13$
Gy; both models fit cluster abundance with no tilt, i.e. $n_p=1$.
From Ref.~\protect\cite{UCLA96}.)
}
\end{figure}

CHDM cosmological models have $\Omega=1$ mostly in cold dark matter
but with a small admixture of hot dark matter, light neutrinos 
contributing $\Omega_\nu = m_{\nu,tot}/(92 h^2 {\rm eV}) \approx 0.2$,
corresponding to a total neutrino mass of $m_{\nu, tot} \approx 5$ eV
for $h=0.5$.  CHDM models are a good fit to much observational data
\cite{PHKC95,CHDM02} --- for example, correlations of galaxies and
clusters and direct measurements of the power spectrum $P(k)$,
velocities on small and large scales, and other statistics such as the
Void Probability Function (probability $P_0(r)$ of finding no bright
galaxy in a randomly placed sphere of radius $r$). My colleagues and I
had earlier shown that CHDM with $\Omega_\nu=0.3$ predicts a VPF
larger than observations indicate~\cite{Ghigna94}, but new results
based on our $\Omega_\nu=0.2$ simulations in which the neutrino mass
is shared equally between two neutrino species~\cite{PHKC95} show that
the VPF for this model is in excellent agreement with observations.
However, our simulations \cite{KPH96} of COBE-normalized \lcdm\  with
$h=0.7$ and $\Omega_0=0.3$ lead to a VPF which is too large to be
compatible with the data \cite{Ghigna96}.

Moreover, there is mounting astrophysical and laboratory data
suggesting that neutrinos have non-zero mass~\cite{PHKC95,FPQ}.  The
analysis of the LSND data through 1995~\cite{LSND96} strengthens the
earlier LSND signal for $\bar\nu_\mu \rightarrow \bar\nu_e$
oscillations. Comparison with exclusion plots from other experiments
implies a lower limit $\Delta m^2_{\mu e} \equiv
|m(\nu_\mu)^2-m(\nu_e)^2| \gsim 0.2$ eV$^2$, implying in turn a lower
limit $m_\nu \gsim 0.45$ eV, or $\Omega_\nu \gsim 0.02 (0.5/h)^2$.
This implies that the contribution of hot dark matter to the
cosmological density is larger than that of all the visible stars
($\Omega_\ast \approx 0.004$~\cite{Peebles}). More data and analysis
are needed from LSND's $\nu_\mu \rightarrow \nu_e$ channel before the
initial hint \cite{Cald95} that $\Delta m^2_{\mu e} \approx 6$ eV$^2$
can be confirmed.  Fortunately the KARMEN experiment has just added
shielding to decrease its background so that it can probe the same
region of $\Delta m^2_{\mu e}$ and mixing angle, with sensitivity as
great as LSND's within about two years.  The Kamiokande
data~\cite{KamAtmNu} showing that the deficit of $E > 1.3$ GeV
atmospheric muon neutrinos increases with zenith angle suggests that
$\nu_\mu \rightarrow \nu_\tau$ oscillations \cite{Whymutau} occur with
an oscillation length comparable to the height of the atmosphere,
implying that $\Delta m^2_{\tau \mu} \sim 10^{-2}$
eV$^2$~\cite{KamAtmNu} --- which in turn implies that if either
$\nu_\mu$ or $\nu_\tau$ have large enough mass ($\gsim 1$ eV) to be a
hot dark matter particle, then they must be nearly degenerate in mass,
i.e. the hot dark matter mass is shared between these two neutrino
species. The much larger Super-Kamiokande detector is now operating,
and we should know by about the end of 1996 whether the Kamiokande
atmospheric neutrino data that suggested $\nu_\mu \rightarrow
\nu_\tau$ oscillations will be confirmed and extended \cite{SKam}.
Starting in 1997 there will be a long-baseline neutrino oscillation
disappearance experiment to look for $\nu_\mu \rightarrow \nu_\tau$
with a beam of $\nu_\mu$ from the KEK accelerator directed at the
Super-Kamiokande detector, with more powerful Fermilab-Soudan,
KEK-Super-Kamiokande, and possibly CERN-Gran Sasso long-baseline
experiments in subsequent years.

Evidence for non-zero neutrino mass evidently favors CHDM, but it also
disfavors low-$\Omega$ models.  Because free streaming of the
neutrinos damps small-scale fluctuations, even a little hot dark
matter causes reduced fluctuation power on small scales and requires
substantial cold dark matter to compensate; thus evidence for even 2
eV of neutrino mass favors large $\Omega$ and would be incompatible
with a cold dark matter density $\Omega_c$ as small as 0.3
\cite{PHKC95}. Allowing $\Omega_\nu$ and the tilt to vary, CHDM can
fit observations over a somewhat wider range of values of the Hubble
parameter $h$ than standard or tilted CDM~\cite{Liddle1}.  This is
especially true if the neutrino mass is shared between two or three
neutrino species \cite{PHKC95,Holtz89,HoltzPri93,PS95}, since then the
lower neutrino mass results in a larger free-streaming scale over
which the power is lowered compared to CDM; the result is that the
cluster abundance predicted with $\Omega_\nu \approx 0.2$ and $h
\approx 0.5$ and COBE normalization (corresponding to $\sigma_8 \approx
0.7$) is in reasonable agreement with observations without the need to
tilt the model\cite{Borgani96} and thereby reduce the small-scale
power further.  (In CHDM with a given $\Omega_\nu$ shared between
$N_\nu=2$ or 3 neutrino species, the linear power spectra are
identical on large and small scales to the $N_\nu=1$ case; the only
difference is on the cluster scale, where the power is reduced by
$\sim 20\%$ \cite{Holtz89,PHKC95}.)

Another consequence of the reduced power on small scales is that
structure formation is more recent in CHDM than in \lcdm. This may
conflict with observations of damped Lyman $\alpha$ systems in quasar
spectra, and other observations of protogalaxies at high redshift,
although the available evidence does not yet permit a clear decision
on this (see below). While the original $\Omega_\nu=0.3$ CHDM model
\cite{DSS92,KHPR93} certainly predicts far less neutral hydrogen in
damped Lyman $\alpha$ systems (identified as protogalaxies with
circular velocities $V_c \geq 50\kms$) than is observed \cite{DLAS,KBHP},
lowering the hot fraction to $\Omega_\nu \approx 0.2$ dramatically 
improves this~\cite{KBHP,Ma95}. Also, the evidence from
preliminary data of a fall-off of the amount of neutral hydrogen in
damped Lyman $\alpha$ systems for $z \gsim 3$~\cite{Storrie} is in
accord with predictions of CHDM~\cite{KBHP}.

However, as for all $\Omega=1$ models, $h \gsim 0.55$
implies $t_0 \lsim 12$ Gyr, which conflicts with age estimates from
globular cluster \cite{Chaboyer96} and white dwarf cooling
\cite{Wood}. The only way to accommodate both large $h$ and large
$t_0$ within the standard FRW framework of General Relativity is to
introduce a positive cosmological constant ($\Lambda>0)$ 
\cite{LLPR,CaPT}.

\lcdm\  flat cosmological models with $\Omega_0 = 1 - \Omega_\Lambda 
\approx 0.3$, where $\Omega_\Lambda \equiv \Lambda/(3H_0^2)$, 
were discussed as an alternative to $\Omega=1$ CDM since the beginning
of CDM \cite{BFPR,Peeb84}.  They have been advocated more recently
\cite{LCDM} both because they can solve the $H_0-t_0$ problem and
because they predict a larger fraction of baryons in galaxy clusters
than $\Omega=1$ models. Early galaxy formation also is often considered to
be a desirable feature of these models. But early galaxy formation
implies that fluctuations on scales of a few Mpc spent more
time in the nonlinear regime, as compared with CHDM models. As has
been known for a long time, this results in excessive clustering on
small scales. My colleagues and I have found that a typical
$\Lambda$CDM model with $h=0.7$ and $\Omega_0=0.3$, normalized to COBE
on large scales (this fixes $\sigma_8\approx 1.1$ for this model), is
compatible with the number-density of galaxy clusters\cite{Borgani96},
but predicts a power spectrum of galaxy clustering in real space that
is much too high for wavenumbers $k=(0.4-1)h/{\rm Mpc}$~\cite{KPH96}.
This conclusion holds if we assume either that galaxies trace the dark
matter, or just that a region with higher density produces more
galaxies than a region with lower density. One can see immediately
from Figure~1 that there will be a problem with this \lcdm\ model,
since the APM power spectrum is approximately equal to the linear
power spectrum at wavenumber $k \approx 0.6 h$ Mpc$^{-1}$, so there is
no room for the extra power that nonlinear evolution certainly
produces on this scale (see Figure~1 of Ref.~\cite{KPH96} and further
discussion below).  The only way to reconcile the model with the
observed power spectrum is to assume that some mechanism causes strong
anti-biasing --- i.e., that regions with high dark matter density
produce fewer galaxies than regions with low density. While
theoretically possible, this seems very unlikely; biasing rather than
anti-biasing is expected, especially on small scales \cite{KNS96}.
Numerical hydro+N-body simulations that incorporate effects of UV
radiation, star formation, and supernovae explosions~\cite{YepesKKK}
do not show any antibias of luminous matter relative to the dark
matter.

Our motivation to investigate this particular \lcdm\  model was to
have $H_0$ as large as might possibly be allowed in the \lcdm\ class
of models, which in turn forces $\Omega_0$ to be rather small in order
to have $t_0 \gsim 13$ Gyr. There is little room to lower the
normalization of this \lcdm\  model by tilting the primordial power
spectrum $P_p(k)=A k^{n_p}$ (i.e., assuming $n_p$ significantly
smaller than the ``Zel'dovich'' value $n_p=1$), since then the fit to
data on intermediate scales will be unacceptable --- e.g., the number
density of clusters will be too small~\cite{KPH96}.  Tilted \lcdm\
models with higher $\Omega_0$, and therefore lower $H_0$ for $t_0
\gsim 13$ Gyr, appear to have a better hope of fitting the available
data, based on comparing quasi-linear calculations to the
data~\cite{KPH96,LiddleLCDM}. But all cosmological models with a
cosmological constant $\Lambda$ large enough to help significantly
with the $H_0-t_0$ problem are in trouble with new observations providing
strong upper limits on $\Lambda$ \cite{Primack96}: gravitational
lensing~\cite{Kochanek}, HST number counts of ellptical
galaxies~\cite{Driver}, and especially the preliminary results from
measurements using high-redshift Type Ia supernovae~\cite{Perl96}. The
analysis of the data from the first 7 of the Type Ia supernovae
from the LBL group \cite{Perl96b} gave $\Omega_0=1-\Omega_\Lambda=
0.94^{+0.34}_{-0.28}$, or equivalently $\Omega_\Lambda=
0.06^{+0.28}_{-0.34}$ ($<0.51$ at the 95\% confidence level).

It is instructive to compare the $\Omega_0=0.3$, $h=0.7$ \lcdm\  model
that we have been considering with standard CDM and with CHDM.  At
$k=0.5 h$ Mpc$^{-1}$, Figs.~5 and 6 of Ref.~\cite{KNP96} show that the
$\Omega_\nu=0.3$ CHDM spectrum and that of a biased CDM model with the
same $\sigma_8=0.67$ are both in good agreement with the values
indicated for the power spectrum $P(k)$ by the APM and CfA data, while
the CDM spectrum with $\sigma_8=1$ is higher by about a factor of two.
As Figure~2 shows, CHDM with $\Omega_\nu=0.2$ in two neutrino
species~\cite{PHKC95} also gives nonlinear $P(k)$ consistent with the
APM data.

\begin{figure}[htb]    
\centering
\centerline{\psfig{file=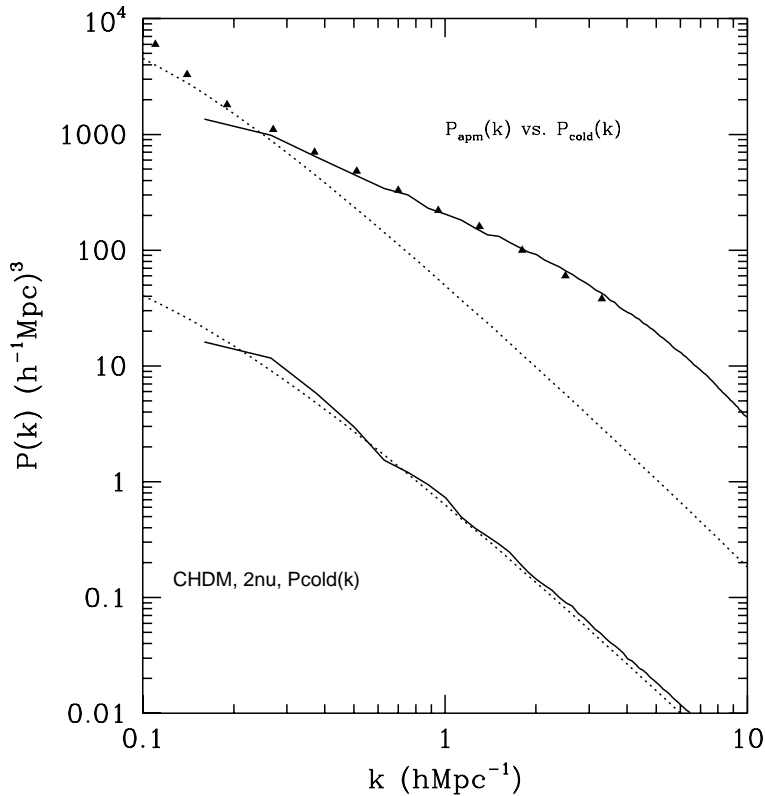,width=10cm}}
\vskip-1pc
\caption{%
Comparison of APM galaxy power spectrum (triangles) with nonlinear
cold particle power spectrum from CHDM model considered in this paper
(upper solid curve).  The dotted curves are linear theory; upper
curves are for $z=0$, lower curves correspond to the higher redshift
$z=9.9$. (From Ref.~\protect\cite{UCLA96}.)}
\end{figure}

\section{Cluster Baryons}

I have recently reviewed the astrophysical data bearing on the values
of the fundamental cosmological parameters, especially $\Omega_0$
\cite{Primack96}.  One of the arguments against $\Omega=1$ that seemed
hardest to answer was the ``cluster baryon crisis''~\cite{WhiteFrenk}:
for the Coma cluster the baryon fraction within the Abell radius is
\begin{equation}
f_b \equiv {M_b \over M_{tot}} \geq 0.009+0.050h^{-3/2},
\end{equation}
where the first term comes from the galaxies and the second from gas.
If clusters are a fair sample of both baryons and dark matter, as they
are expected to be based on simulations, then this is 2-3 times the
amount of baryonic mass expected on the basis of BBN in an $\Omega=1$,
$h\approx 0.5$ universe, though it is just what one would expect in a
universe with $\Omega_0 \approx 0.3$. The fair sample hypothesis
implies that
\begin{equation}
\Omega_0 = {\Omega_b \over f_b}
         = 0.33 \left({\Omega_b \over 0.05}\right)
           \left({0.15 \over f_b}\right).
\end{equation}

A review of the quantity of X-ray emitting gas in a sample of
clusters~\cite{Fabian} finds that the baryon mass fraction within
about 1 Mpc lies between 10 and 22\% (for $h=0.5$; the limits scale as
$h^{-3/2}$), and argues that it is unlikely that (a) the gas could be
clumped enough to lead to significant overestimates of the total gas
mass --- the main escape route considered in \cite{WhiteFrenk} (cf.
also~\cite{ClusBar}). If $\Omega=1$, the alternatives are then either
(b) that clusters have more mass than virial estimates based on the
cluster galaxy velocities or estimates based on hydrostatic
equilibrium \cite{BBlanchard} of the gas at the measured X-ray
temperature (which is surprising since they agree~\cite{BLubin}),
(c) that the usual BBN estimate $\Omega_b \approx 0.05 (0.5/h)^2$ is
wrong, or (d) that the fair sample hypothesis is wrong
\cite{MushotzkyL}.  Regarding (b), it is interesting that there are 
indications from weak
lensing~\cite{Kaiser} that at least some clusters may actually have
extended halos of dark matter --- something that is expected to a
greater extent if the dark matter is a mixture of cold and hot
components, since the hot component clusters less than the
cold~\cite{BHNPK,KofmanKP}. If so, the number density of clusters as a
function of mass is higher than usually estimated, which has
interesting cosmological implications (e.g., $\sigma_8$ is a little
higher than usually estimated).  It is of course possible that the
solution is some combination of alternatives (a)-(d). If
none of the alternatives is right, then the only conclusion left is
that $\Omega_0 \approx 0.33$.  The cluster baryon problem is clearly
an issue that deserves very careful examination.

It has recently been argued \cite{SSchramm} that CHDM models are
compatible with the X-ray data within observational uncertanties of
both the BBN predictions and X-ray data. Indeed, the rather high
baryon fraction $\Omega_b \approx 0.1 (0.5/h)^2$ implied by recent
measurements of low D/H in two high-redshift Lyman limit
systems~\cite{Tytler,Hogan} helps resolve the cluster baryon crisis
for all $\Omega=1$ models --- it is escape route (c) above.  With the
higher $\Omega_b$ implied by the low D/H, there is now a ``baryon
cluster crisis'' for low-$\Omega_0$ models!  Even with a baryon
fraction at the high end of observations, $f_b \lsim 0.2
(h/0.5)^{-3/2}$, the fair sample hypothesis with this $\Omega_b$
implies $\Omega_0 \gsim 0.5 (h/0.5)^{-1/2}$.

\section{Warm Dark Matter vs. CHDM}

It will be instructive to look briefly at Warm Dark Matter (WDM), both
to see that some variants of CDM have less success than others in
fitting cosmological observations, and also because there is renewed
interest in WDM.  Although CHDM and WDM are similar in the sense that
both are intermediate models between CDM and HDM, CHDM and WDM are
quite different in their implications.

The problems with a pure Hot Dark Matter (HDM) adiabatic cosmology are
well known: free-streaming of the hot dark matter completely destroys
small-scale fluctuations, so that the first structures that can form
are on the mass scale of clusters or superclusters, and galaxies must
form by fragmentation of these larger structures; but observations
show that galaxies are much older than superclusters, which have low
overdensity and are still forming. Moreover, with the COBE upper limit
to the normalization of HDF, hardly any structure at all will form
even by the present epoch.

WDM is a simple modification of HDM, obtained by changing the assumed
average number density $n$ of the particles.  In the usual HDM, the
dark matter particles are neutrinos, each species of which has
$n_\nu=108$ cm$^{-3}$, with a corresponding mass of $m_0 =
\Omega_\nu \rho_0/n_\nu = \Omega_\nu 92 h^2$ eV, with $\Omega_\nu = 
1- \Omega_b \approx 1$.  In WDM, there is
a new parameter, $m/m_0$, the ratio of the mass of the warm particle
to the above neutrino mass; correspondingly, the number density of the
warm particles is reduced by the inverse of this factor, so that their
total contribution to the cosmological density is unchanged. Pagels
and I \cite{PP82} proposed perhaps the first WDM particle
candidate, a light gravitino, which was the lightest supersymmetric
particle (LSP) in the earliest version of supersymmetric phenomenology
(which was subsequently largely abandoned in favor of a hidden-sector
sypersymmetry breaking scheme, but which is now being reconsidered:
cf. \cite{Dine96}). Olive \& Turner \cite{OT82} proposed left-handed
neutrinos as WDM.  In both cases, these particles interact much more
weakly than neutrinos, decouple earlier from the hot big bang, and
thus have diluted number density compared to neutrinos since they do
not share in the entropy released by the subsequent annihilation of
species such as quarks. This is analogous to (but more extreme than)
the neutrinos themselves, which have lower number density today than
photons because the neutrinos decouple before e$^+$e$^-$ annihilation
(and also because they are fermions).

In order to investigate the cosmological implications of any dark
matter candidate, it is necessary to work out the gravitational
clustering of these particles, first in linear theory, and then after
the amplitude of the fluctuations grows into the nonlinear regime.
Colombi, Dodelson, \& Widrow \cite{WDM} recently did this for WDM, and
Figure 1 in their paper compares the square of the linear transfer
functions for WDM and CHDM.

One often can study large scale structure just on the basis of such
linear calculations, without the need to do computationally expensive
simulations of the non-linear gravitational clustering.  Such studies
have shown that matching the observed cluster and galaxy correlations
on scales of about 20-30 $h^{-1}$ Mpc in CDM-type theories requires
that the ``Excess Power'' $EP\approx 1.3$, where
\begin{equation}
EP \equiv {{\sigma(25 \hMpc)/\sigma(8 \hMpc)} \over
              {[\sigma(25 \hMpc)/\sigma(8 \hMpc)]_{sCDM}}},
\end{equation}
and as usual $\sigma(r)=(\delta \rho/\rho)(r)$ is the rms fluctuation
amplitude in randomly placed spheres of radius $r$. The $EP$ parameter
was introduced in the COBE-DMR interpretation paper
\cite{Wright92}, and in a recent paper \cite{Borgani96} we have
shown that $EP$ is related to the spectrum shape parameter $\Gamma$
\cite{EfBW92}
by $\Gamma \approx 0.5 (EP)^{-3.3}$. For CHDM and other models,
this is a useful generalization since the cluster correlations do seem to
be a function of this generalized $\Gamma$, with $\Gamma \approx 0.23$
to match cluster correlation data  \cite{Borgani96}. Peacock \& Dodds
\cite{PD94} have shown that $\Gamma \approx 0.23$ also is required to match
large scale galaxy clustering data.

Since calculating $\sigma(r)$ is a simple matter of integrating the power
spectrum times the top-hat window function,
\begin{equation}
\sigma(r)= \int_0^\infty P(k) W(kr) k^2 dk
\end{equation}
the linear calculations immediately allow determination of $EP$ for
WDM and CHDM.  Ref.~\cite{WDM} shows that for WDM to give the required
$EP$, the parameter value $m/m_0 \approx 1.5-2$, while for CHDM the
required value of the CHDM parameter is $\Omega_\nu \approx 0.3$. But
for WDM with $m/m_0 \gsim 2$, the spectrum lies a lot lower than the
CDM spectrum at $k\gsim 0.3 h^{-1}$ Mpc (length scales $\lambda \lsim
20 \hMpc$), which in turn implies that formation of galaxies,
corresponding to the gravitational collapse of material in a region of
size $\sim 1$ Mpc, will be strongly suppressed compared to CDM.  Thus
WDM will not be able to accommodate simultaneously the distribution of
clusters and galaxies. But CHDM will do much better.

Probably the only way to accommodate WDM in a viable cosmological
model is as part of a mixture with hot dark matter, which might even
arise naturally in a supersymmetric model \cite{BorWDM} of the sort in
which the gravitino is the LSP \cite{Dine96}.  Cold plus ``volatile''
dark matter is a related possibility \cite{BonoVDM}.

There are many more parameters needed to describe the presently
available data on the distribution of galaxies and clusters and their
formation history than the few parameters needed to specify a CDM-type
model. Thus it should not be surprising that at most a few CDM variant
theories can fit all this data.  Once it began to become clear that
standard CDM was likely to have problems accounting for all the data,
after the discovery of large-scale flows of galaxies was announced in
early 1986 \cite{Burstein86}, I advised Jon Holtzman in his
dissertation research to work out the linear theory for a wide variety
of CDM variants \cite{Holtz89} so that we could see which ones would
best fit the data \cite{HoltzPri93}.  The clear winners were CHDM with
$\Omega_\nu\approx 0.3$ if $h\approx 0.5$, and \lcdm\ with $\Omega_0
\approx 0.2$ if $h\approx 1$.  Variants of both these models remain
perhaps the best bets still.

\section{CHDM: Early Structure Troubles?}

Aside from the possibility mentioned at the outset that the Hubble
constant is too large and the universe too old for any $\Omega=1$
model to be viable, the main potential problem for CHDM appears to be
forming enough structure at high redshift. Although, as I mentioned
above, the prediction of CHDM that the amount of gas in damped Lyman
$\alpha$ systems is starting to decrease at high redshift $z \gsim 3$
seems to be in accord with the available data, the large velocity
spread of the associated metal-line systems {\it may} indicate that
these systems are more massive than CHDM would predict (see
e.g.,~\cite{Lu,Wolfe}).  Also, results from a recent CDM hydrodynamic
simulation \cite{HernDLAS} in which the amount of neutral hydrogen in
protogalaxies seemed consistent with that observed in damped Lyman
$\alpha$ systems led the authors to speculate that CHDM models would
produce less than enough; however, since the regions identified as
damped Lyman $\alpha$ systems in the simulations were not actually
resolved, this will need to be addressed by higher resolution
simulations for all the models considered.

Finally, Steidel et al.~\cite{Steidel} have found objects by their
emitted light at redshifts $z=3-3.5$ apparently with relatively high
velocity dispersions (indicated by the equivalent widths of absorption
lines), which they tentatively identify as the progenitors of giant
elliptical galaxies. {\it Assuming} that the indicated velocity
dispersions are indeed gravitational velocities, Mo \& Fukugita
(MF)~\cite{MoF} have argued that the abundance of these objects is
higher than expected for the COBE-normalized $\Omega=1$ CDM-type
models that can fit the low-redshift data, including CHDM, but in
accord with predictions of the \lcdm\ model considered here. (In more
detail, the MF analysis disfavors CHDM with $h=0.5$ and $\Omega_\nu
\gsim 0.2$ in a single species of neutrinos. They apparently would
argue that this model is then in difficulty since it overproduces rich
clusters --- and if that problem were solved with a little tilt $n_p
\approx 0.9$, the resulting decrease in fluctuation power on small
scales would not lead to formation of enough early objects. However,
if $\Omega_\nu \approx 0.2$ is shared between two species of
neutrinos, the resulting model appears to be at least marginally
consistent with both clusters and the Steidel objects even with the
assumptions of MF.  The \lcdm\ model with $h=0.7$ consistent with the
most restrictive MF assumptions has $\Omega_0 \gsim 0.5$, hence $t_0
\lsim 12$ Gyr.  \lcdm\ models having tilt and lower $h$, and therefore
more consistent with the small-scale power constraint discussed above,
may also be in trouble with the MF analysis.) But in addition to
uncertainties about the actual velocity dispersion and physical size
of the Steidel et al. objects, the conclusions of the MF analysis can
also be significantly weakened if the gravitational velocities of the
observed baryons are systematically higher than the gravitational
velocities in the surrounding dark matter halos, as is perhaps the
case at low redshift for large spiral galaxies~\cite{NFW}, and even
more so for elliptical galaxies which are largely self-gravitating
stellar systems in their central regions.

Given the irregular morphologies of the high-redshift objects seen in
the Hubble Deep Field \cite{vdB96} and other deep HST images, it seems
more likely that they are relatively low mass objects undergoing
starbursts, possibly triggered by mergers,
rather than galactic protospheroids.  Since the number density of the
brightest of such objects may be more a function of the probability
and duration of such starbursts rather than the nature of the
underlying cosmological model, it may be more useful to use the star
formation or metal injection rates \cite{Madau} indicated by the total
observed rest-frame ultraviolet light to constrain models
\cite{SomPri}.  The available data on the history of star formation
\cite{Gallego,Lilly,Madau} suggests that most of the stars and most of
the metals observed formed relatively recently, after about redshift
$z\sim 1$; and that the total star formation rate at $z\sim 3$ is
perhaps a factor of 3 lower than at $z \sim 3$, with yet another
factor  of $\sim 3$ falloff to $z \sim 4$ (although the rates at
$z\gsim 3$ could be higher if most of the star formation is in objects
too faint to see). This is in accord with
indications from damped Lyman $\alpha$ systems \cite{FallCP} and
expectations for $\Omega=1$ models such as CHDM, but not with the
expectations for low-$\Omega_0$ models which have less growth of
fluctuations at recent epochs, and therefore must form structure
earlier.  But this must be investigated using more detailed modelling,
including gas cooling and feedback from stars and supernovae
\cite{SomPri}, before strong conclusions can be drawn.

There is another sort of constraint from observed numbers of
high-redshift protogalaxies that would appear to disfavor \lcdm. The
upper limit on the number of $z\gsim 4$ objects in the Hubble Deep
Field (which presumably correspond to smaller-mass galaxies than most
of the Steidel objects) is far lower than the expectations in
low-$\Omega_0$ models, especially with a positive cosmological
constant, because of the large volume at high redshift in such
cosmologies~\cite{Yahil}. Thus evidence from high-redshift objects
cuts both ways, and it is too early to tell whether high- or
low-$\Omega_0$ models will ultimately be favored.

\section{Advantages of Mixed CHDM Over Pure CDM Models}

There are three basic reasons why a mixture of cold plus hot dark
matter works better than pure CDM without any hot particles: {\bf (1)}
the power spectrum shape $P(k)$ is a better fit to observations, {\bf
(2)} there are indications from observations for a more weakly
clustering component of dark matter, and {\bf (3)} a hot component may
help avoid the too-dense central dark matter density in pure CDM dark
matter halos. I will discuss each in turn.

{\bf (1) Spectrum shape.}  As I explained in discussing WDM vs. CHDM
above, the pure CDM spectrum $P(k)$ does not fall fast enough on the
large-$k$ side of its peak in order to fit indications from galaxy and
cluster correlations and power spectra.  The discussion there of
``Excess Power'' is a way of quantifying this.  This is also related
to the overproduction of clusters in pure CDM. The obvious way to
prevent $\Omega=1$ sCDM normalized to COBE from overproducing clusters
is to tilt it a lot (the precise amount depending on how much of the
COBE fluctuations are attributed to gravity waves, which can be
increasingly important as the tilt is increased).  But a constraint on
CDM-type models that is likely to follow both from the high-$z$ data
just discussed and from the preliminary indications on cosmic
microwave anisotropies at and beyond the first acoustic peak from the
Saskatoon experiment \cite{Netterfield} is that viable models cannot
have much tilt, since that would reduce too much both their
small-scale power and the amount of small-angle CMB anisotropy. As I
have already explained, by reducing the fluctuation power on cluster
scales and below, COBE-normalized CHDM naturally fits both the CMB
data and the cluster abundance without requiring much tilt.  The need
for tilt is further reduced if a high baryon fraction $\Omega_b
\gsim 0.1$ is assumed \cite{LiddleHib}, 
and this also boosts the predicted height of the first acoustic peak. 
No tilt is necessary for
$\Omega_\nu=0.2$ shared between $N_\nu=2$ neutrino species with $h=0.5$
and $\Omega_b=0.1$. Increasing the Hubble parameter in COBE-normalized
models increases the amount of small-scale power, so that if we raise
the Hubble parameter to $h=0.6$ keeping $\Omega_\nu=0.2$ and
$\Omega_b=0.1(0.5/h)^2=0.069$, then fitting the cluster abundance in
this $N_\nu=2$ model requires tilt $1-n_p \approx 0.1$ with no gravity waves
(i.e., $T/S=0$; alternatively if $T/S=7(1-n_p)$ is assumed, about half
as much tilt is needed, but the observational consequences are mostly
very similar, with a little more small scale power). The fit to the 
small-angle CMB data is still good, and
the predicted $\Omega_{\rm gas}$ in damped Lyman $\alpha$ systems is a
little higher than for the $h=0.5$ case. The only obvious problem with
$h=0.6$ applies to any $\Omega=1$ model --- the universe is rather
young: $t_0=10.8$ Gyr.

{\bf (2) Need for a less-clustered component of dark matter.}  The
fact that group and cluster mass estimates on scales of $\sim 1
\hMpc$ typically give values for $\Omega$ around 0.1-0.2 \cite{NBahcall},
while larger-scale estimates give larger values around 0.3-1
\cite{Dekel94} suggests that there is a component of dark matter that
does not cluster on small scales as efficiently as cold dark matter is
expected to do.  In order to  quantify this, my colleagues and I have
performed the usual group $M/L$ measurement of $\Omega_0$ on small
scales in ``observed'' $\Omega=1$ simulations of both CDM and CHDM
\cite{NKP96}. We found that COBE-normalized $\Omega_\nu=0.3$ CHDM
gives $\Omega_{M/L} = 0.12-0.18$ compared to $\Omega_{M/L} = 0.15$ for
the CfA1 catalog analyzed exactly the same way, while for CDM
$\Omega_{M/L} = 0.34-0.37$, with the lower value corresponding to bias
$b=1.5$ and the higher value to $b=1$ (still below the COBE
normalization).  Thus local measurements
of the density in $\Omega=1$ simulations can give low values, but it
helps to have a hot component to get values as low as observations
indicate.  We found that there are three reasons why this virial
estimate of the mass in groups misses so much of the matter in the
simulations: (1) only the mass within the mean harmonic radius $r_h$
is measured by the virial estimate, but the dark matter halos of
groups continue their roughly isothermal falloff to at least $2r_h$,
increasing the total mass by about a factor of 3 in the CHDM
simulations; (2) the velocities of the galaxies are biased by about
$70\%$ compared to the dark matter particles, which means that the
true mass is higher by about another factor of 2; and (3) the groups
typically lie along filaments and are significantly elongated, so the
spherical virial estimator misses perhaps 30\% of the mass for this
reason.  Our visualizations of these simulations \cite{BHNPK} show
clearly how extended the hot dark matter halos are.  An analysis of
clusters in CHDM found similar effects, and suggested that
observations of the velocity distributions of galaxies around clusters
might be able to discriminate between pure cold and mixed cold + hot
models \cite{KofmanKP}.  This is an area where more work needs to be
done --- but it will not be easy since it will probably be necessary
to include stellar and supernova feedback in identifying galaxies in
simulations, and to account properly for foreground and background
galaxies in observations.

{\bf (3) Preventing too dense centers of dark matter halos.} Flores
and I \cite{FP94} pointed out that dark matter density profiles with
$\rho(r) \propto r^{-1}$ near the origin from high-resolution
dissipationless CDM simulations \cite{CDMsims} are in serious conflict
with data on dwarf spiral galaxies (cf. also Ref. \cite{Moore}), and
in possible conflict with data on larger spirals \cite{FPBF93} and on
clusters (cf. \cite{Miralda,FP96}). Navarro, Frenk, \& White
\cite{NFW} agree that rotation curves of small spiral galaxies such as
DDO154 and DDO170 are strongly inconsistent with their universal dark
matter profile $\rho_{NFW}(r) \propto 1/[r(r+a)^2]$.  I am at present
working with Stephane Courteau, Sandra Faber, Ricardo Flores, and
others to see whether $\rho_{NFW}$ is consistent with data from high-
and low-surface-brightness galaxies with moderate to large circular
velocities are consistent with this universal profile. The failure of
simulations to form cores as observed in dwarf spiral galaxies either is a
clue to a property of dark matter that we don't understand, or is
telling us the simulations are inadequate. It is important to discover
whether this is a serious problem, and whether inclusion of hot dark
matter or of dissipation in the baryonic component of galaxies can
resolve it.  It is clear that including hot dark matter will decrease
the central density of dark matter halos, both because the lower
fluctuation power on small scales in such models will prevent the
early collapse that produces the highest dark matter densities, and
also because the hot particles cannot reach high densities because of
the phase space constraint \cite{TremaineGunn,KofmanKP}.  But this may
not be enough.

\section{Best Bet CDM-Type Models}

As I said at the outset, I think CHDM is the best bet if $\Omega_0$
turns out to be near unity and the Hubble parameter is not too large,
while \lcdm\ is the best bet if the Hubble parameter is too large to
permit the universe to be older than its stars with $\Omega=1$.

Both theories do seem less ``natural'' than sCDM.  But although sCDM
won the beauty contest, it doesn't fit the data.  CHDM is just sCDM
with some light neutrinos.  After all, we know that neutrinos exist,
and there is experimental evidence --- admittedly not yet entirely
convincing --- that at least some of these neutrinos have mass,
possibly in the few-eV range necessary for CHDM.

Isn't it an unnatural coincidence to have three different sorts of
matter --- cold, hot, and baryonic --- with contributions to the
cosmological density that are within an order of magnitude of each
other?  Not necessarily.  All of these varieties of matter may have
acquired their mass from (super?)symmetry breaking associated with the
electroweak phase transition, and when we understand the nature of the
physics that determines the masses and charges that are just
adjustable parameters in the Standard Model of particle physics, we
may also understand why $\Omega_c$, $\Omega_\nu$, and $\Omega_b$ are so
close.  In any case, CHDM is certainly not uglier than \lcdm.

In the \lcdm\ class of models, the problem of too much power on small
scales that I discussed at some length for $\Omega_0=0.3$ and $h=0.7$
\lcdm\ implies either that there must be some physical mechanism that
produces strong, scale-dependent anti-biasing of the galaxies with
respect to the dark matter, or else that higher $\Omega_0$ and lower
$h$ are preferred, with a significant amount of tilt to get the
cluster abundance right and avoid too much small-scale power
\cite{KPH96}.  Higher $\Omega_0 \gsim 0.5$ also is more consistent with the
evidence summarized above against large $\Omega_\Lambda$ and in favor
of larger $\Omega_0$, especially in models such as \lcdm\ with
Gaussian primordial fluctuations.  But then $h\lsim0.63$ for $t_0
\gsim 13$ Gyr.

Among CHDM models, having $N_\nu=2$ species share the neutrino mass
gives a better fit to COBE, clusters, and small-scall data than
$N_\nu=1$, and moreover it appears to be favored by the available
experimental data \cite{PHKC95}.  But it remains to be seen whether
CHDM models can fit the data on structure formation at high redshifts.

\bigskip

\noindent {\bf Acknowledgments.}  This work was partially supported
by NASA and NSF grants at UCSC. I thank my collaborators, especially
Anatoly Klypin, for many helpful discussions of the material presented
here.

\def\mnras{MNRAS}
\def\aa{A\&A}
\def\apj{ApJ}
\def\apjs{ApJS}


\begin{thebibliography}{00}
\bibitem{Whymutau} The Kamiokande data is consistent with
atmospheric $\nu_\mu$ oscillating to any other neutrino species with a
large mixing angle. But (see discussion and references in, e.g.,
\cite{PHKC95,FPQ}) $\nu_\mu$ oscillating to $\nu_e$ with a large
mixing angle is probably inconsistent with reactor and other data, and
$\nu_\mu$ oscillating to a sterile neutrino $\nu_s$ (i.e., one that
does not interact via the usual weak interactions) with a large mixing
angle is inconsistent with the usual Big Bang Nucleosynthesis
constraints.
\end{thebibliography}
\end{document}